\documentclass[fleqn,twoside,twocolumn]{article}
\usepackage{espcrc2}
\usepackage{amsmath}

\usepackage{graphicx}
\usepackage[figuresright]{rotating}

\usepackage{flushend, cuted, widetext}

\newenvironment{packed_itemize}{
\begin{itemize}
  \setlength{\itemsep}{1pt}
  \setlength{\parskip}{0pt}
  \setlength{\parsep}{0pt}
}{\end{itemize}}

\include{symbols}

\title{The \BdbKsmm decay}

\author{U. Egede\address[Imperial]{Imperial College London, London
    SW7~2AZ, United Kingdom}
  T. Hurth\address{Institute for Physics, Johannes
    Gutenberg-University, D-55099 Mainz, Germany}
  J. Matias\address[IFAE]{Universitat Aut\`onoma de Barcelona, 08193
    Bellaterra, Barcelona, Spain}
  M. Ramon\addressmark[IFAE]
  W. Reece\address{CERN, Dept. of Physics, CH-1211 Geneva 23, Switzerland}}
\begin{document}

\begin{abstract}
  \noindent In this paper the potential for the discovery of new
  physics in the exclusive decay \BdbKsmm is discussed. Attention is
  paid to constructing observables which are protected from
  uncertainties in QCD form factors and at the same time observe the
  symmetries of the angular distribution. We discuss the sensitivity
  to new physics in the observables including the effect of
  \CP-violating phases.  \vspace{1pc}
\end{abstract}

\maketitle

\section{Introduction}
\vspace{-15cm}
\hspace{14cm}\mbox{MZ-TH/10-41}
\vspace{14.5cm}

With the \lhcb experiment coming online, there is the prospects of
performing precision physics in the \BdbKsmm channel within a few years. This
means particular attention has to be given to how the predictions from
the phenomenology and the experimental measurements are compared.

First published results
from \belle~\cite{:2009zv} and \babar~\cite{Aubert:2006vb} based on
$\order(100)$ decays already demonstrate their feasibility.

In~\cite{Kruger:2005ep}, it was proposed  to  construct observables that
maximise the sensitivity to contributions driven by the
electro-magnetic dipole operator \Opep7, while, at the same time,
minimising the dependence on the poorly  known  soft form factors.

\AT2 is highly sensitive to new right-handed currents driven by
the  operator \Opep7~\cite{Lunghi:2006hc},  to which \AFB
is blind.

Looking for the complete set of angular observables sensitive to
right-handed currents, one is guided to the construction of \AT3 and
\AT4~\cite{Egede:2008uy} and \AT5~\cite{Egede:2010zc}. The observables
\AT{i} (with $i=2,3,4,5$) use the \Kstarz spin amplitudes as the
fundamental building block. This provides more freedom to disentangle
the information on specific Wilson coefficients than just restricting
oneself to use the coefficients of the angular distribution as it was
recently done in \cite{Altmannshofer:2008dz}. For instance, \AT2,
being directly proportional to \Cp7 enhances its sensitivity to the
type of NP entering this coefficient. Moreover using selected ratios
of the coefficients of the distribution, like in \AT{i}, the
sensitivity to soft form factors is completely canceled out at LO.

\section{Differential decay distribution}
\label{sec:distribution}
The decay \BdbKsmm, with $\Kstarzb \to \Km \pip$ on the mass shell, is
completely described by four independent kinematic variables, the
lepton-pair invariant mass squared, \qsq, and the three angles
$\theta_l$, $\theta_{K}$, $\phi$. Summing over the spins of the final
state particles, the differential decay distribution of \BdbKsll can
be written as
\begin{equation}
\label{eq:differential decay rate}
  \frac{d^4\Gamma}{dq^2\, d\cos\theta_l\, d\cos\theta_{K}\, d\phi} =
   \frac{9}{32\pi} J(q^2, \theta_l, \theta_{K}, \phi)\,,
\end{equation}
The dependence on the three angles can be made more explicit:

\begin{widetext}
\begin{eqnarray}
  &&\hspace{-4em} J(q^2, \theta_l, \theta_K, \phi) \nonumber\\
    &=&  J_{1s} \sin^2\theta_K + J_{1c} \cos^2\theta_K
      + (J_{2s} \sin^2\theta_K + J_{2c} \cos^2\theta_K) \cos 2\theta_l + J_3 \sin^2\theta_K \sin^2\theta_l \cos 2\phi 
\nonumber \\       
    && + J_4 \sin 2\theta_K \sin 2\theta_l \cos\phi  + J_5 \sin 2\theta_K \sin\theta_l \cos\phi+ (J_{6s} \sin^2\theta_K +  {J_{6c} \cos^2\theta_K})  \cos\theta_l 
\nonumber \\      
    && + J_7 \sin 2\theta_K \sin\theta_l \sin\phi  + J_8 \sin 2\theta_K \sin 2\theta_l \sin\phi + J_9 \sin^2\theta_K \sin^2\theta_l \sin 2\phi\,.
\label{eq:J}
\end{eqnarray}
\end{widetext}

The $J_i$ depend on products of the six complex $K^*$ spin amplitudes,
$\apaLR$, $\apeLR$ and $\azeLR$ in the case of the SM with massless
leptons. The $L$ and $R$ indicate a left and right handed current
respectively. Each of these is a function of $q^2$. The amplitudes are
just linear combinations of the well-known helicity amplitudes
describing the $B\to K\pi$ transition:
\begin{equation}
  \label{hel:trans}
  A_{\bot,\|}^{L.R} = (H_{+1}^{L.R}\mp H_{-1}^{L.R})/\sqrt{2}\, , 
  \qquad A_0^{L,R}=H_0^{L.R} \, .  
\end{equation}
The $J_i$ will be bi-linear functions of the spin amplitudes such as
\begin{equation}
  \label{eq:J1s}
  J_{1s} = \frac{3}{4} \left[|\apeL|^2 + |\apaL|^2 + |\apeR|^2 + |\apaR|^2  \right],
\end{equation}
with the expression for the eleven other $J_i$ terms given
in~\cite{Egede:2010zc}.

The amplitudes themselves can be parametrised in terms of the seven
$B\to K^*$ form factors by means of a narrow-width approximation. They
also depend on the short-distance Wilson coefficients $\mathcal{C}_i$
corresponding to the various operators of the effective electroweak
Hamiltonian. The precise definitions of the form factors and of the
effective operators are given in~\cite{Egede:2008uy}. Assuming
only the three most important SM operators for this decay mode, namely
\Ope7, \Ope9, and \Ope{10}, and the chirally flipped ones, being
numerically relevant, we have as an example
\begin{align}
    \hspace{-0.6cm} \apeLR = &  N \sqrt{2} \lambda^{1/2}
     \bigg[ \frac{ V(q^2) }{ m_B + m_\kstar} \bigg\{  \nonumber \\
      & \quad (\Ceff9 + \Cpeff9) \mp (\Ceff{10} + \Cpeff{10})
    \bigg\}  +  \nonumber \\
    & + \frac{2m_b}{q^2} (\Ceff7 + \Cpeff7) T_1(q^2) \bigg]
\end{align}
where the $\mathcal{C}_i$ denote the corresponding Wilson
coefficients, $N$ is a normalisation and
\begin{equation}
  \label{eq:Lambdadef}
  \lambda= m_B^4  + m_{K^*}^4 + q^4 - 2 (m_B^2 m_{K^*}^2+ 
  m_{K^*}^2 \qsq  + m_B^2 \qsq).
\end{equation}
There are similar expressions for the other spin
amplitudes~\cite{Egede:2008uy}.

When going from the six complex spin amplitude to the expression of
the angular distribution (\ref{eq:J}) with 12 $J_i$ terms, one would
naturally assume there is no loss of information. However, it turns
out that there are a number of relations between the $J_i$ terms; this
in turns means that there are continuous transformations of the spin
amplitudes that will result in the identical angular distribution. The
full derivation of these symmetries can be found in~\cite{Egede:2010zc},
while here we just give the result.

In total four of the $J_i$ terms can be written as a function of the
eight remaining $J_i$. Thus, the differential distribution is invariant
under the following four independent symmetry transformations of the
amplitudes
\begin{align}
  \label{eq:SymMassless}
  n_i^{'} = &
  \left[
    \begin{array}{ll}
      e^{i\phi_L} & 0 \\
      0 & e^{-i \phi_R}
    \end{array}
  \right]
  \left[
    \begin{array}{rr}
      \cos \theta & -\sin \theta \\
      \sin \theta & \cos \theta
    \end{array}
  \right] \nonumber \\
  & \left[
    \begin{array}{rr}
      \cosh i \tilde{\theta} &  -\sinh i \tilde{\theta} \\
      -\sinh i \tilde{\theta} & \cosh i \tilde{\theta}
    \end{array}
  \right]
  n_i \,,
\end{align}
where $\phi_L$, $\phi_R$, $\theta$ and $\tilde{\theta}$ can be varied
independently and $n_i$ is defined as
\begin{align}
  n_1&=(\apaL, \apaR^*) \,, \nonumber \\
  n_2&=(\apeL, - \apeR^*) \,, \\
  n_3&=(\azeL, \azeR^*) \, \nonumber .
\end{align}

Normally, there is the freedom to pick a single global
phase, but as $L$ and $R$ amplitudes do not interfere here, two phases
can be chosen arbitrarily as reflected in the first transformation
matrix. The interpretation of the third and fourth symmetry is that
they transform a helicity $+1$ final state with a left handed current
into a helicity $-1$ state with a right handed current. As we
experimentally cannot measure the simultaneous change of helicity and
handedness of the current, these transformations turn into symmetries
for the differential decay rate.

\section{Comparing theory and experiment}
As can be seen in the previous section it is possible to express the
full angular dependence in terms of the effective Wilson
coefficients. From this it would seem that it is trivially possible to
extract full knowledge of the Wilson coefficients from a fit to
experimental data on the angular distribution. Unfortunately there are
several problems related to such an approach that will be described in
turn.

As we are dealing with an exclusive decay, we have to rely on the
calculation of form factors. These can come from either light cone sum
rule calculations in the low \qsq region, or from lattice QCD
calculations in the high \qsq region but are in both cases subject to
significant uncertainty. In the low \qsq region below 6\gevgevcccc the
use of soft collinear effective theory~(SCET), allows for the a reduction in
the number of form factors from seven to two, which we will
subsequently take advantage of. From below we are limited to $\qsq >
1\gevgevcccc$ due to a logarithmic divergence in the SCET approach.

Even after the form factors have been considered, each spin amplitude
has an uncertainty due to $\lqcd/m_\b$ corrections. The level of these are
not known but dimensional considerations leads one to expect them at
the 10\% level or below. In our analysis we have made the effect
explicit by illustrating the effect of $\lqcd/m_\b$ corrections at the 5\%
and 10\% level.

The effect of charm loops will be important even outside the narrow
resonance regions of the $\Bd \to \jpsi \Kstarz$ and $\Bd \to \psitwos
\Kstarz$ due to the effect of virtual \ccbar
pairs~\cite{Khodjamirian:2010vf}. The effect is small for
$\qsq<6\gevgevcccc$ which is the region we consider.

When comparing theory to experimental data, two approaches are in
general taken. One can start at the theory end with a physics model;
from that we calculate the Wilson coefficients and subsequently the
spin amplitudes. The last step will lead to a loss of accuracy due to
the form factor uncertainties and the unknown $\lqcd/m_\b$ corrections.
Finally we calculate the angular coefficients $J_i$ which can be
compared directly to an angular fit of experimental data.

The other approach is to start from the experimental determination of
the angular coefficients and from that determine the spin
amplitudes. As we have seen in the previous section this process is
not well defined due to symmetries in the angular
distribution. Ignoring this point one can from the spin amplitudes get
the Wilson coefficients (again suffering from form factor and
$\lqcd/m_\b$ uncertainties) and go onto a physics model.

\begin{figure}
  \centering
  \includegraphics[width=0.98\linewidth]{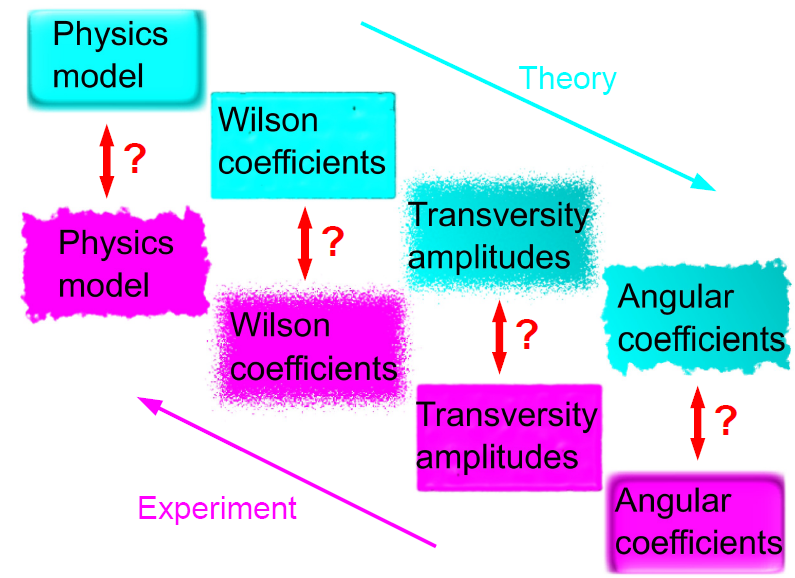}
  \caption{Traditionally experiment and theory are compared at some
    point along the route of transforming one to the other. For
    \BdbKsmm this has the disadvantage of having a large theoretical
    uncertainty from form factor and $\lqcd/m_\b$ uncertainties.}
  \label{fig:Compare}
\end{figure}

\begin{figure}
  \centering
  \includegraphics[width=0.98\linewidth]{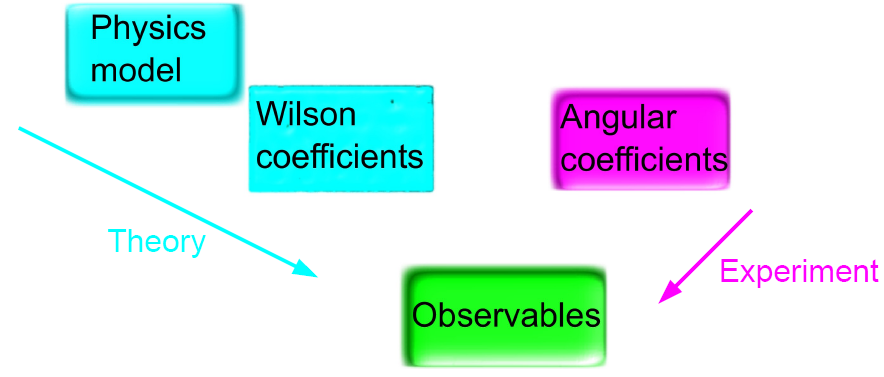}
  \caption{Observables can be constructed which from the theoretical
    side offers cancellation of form factors, and which from the
    experimental side are well defined and have good experimental
    sensitivity.\vspace{-0.5cm}}
  \label{fig:Observable}
\end{figure}

As illustrated in Fig.~\ref{fig:Compare} the experimental results and
the theory can be compared at several different levels. However, the
uncertainties introduced in each direction means this is not
optimal. We suggest instead to create a set of observables which can
be derived from both the theory side and the experimental results as
illustrated in Fig.~\ref{fig:Observable}. In this way the majority of
the uncertainty due to the soft form factors can be eliminated.

\section{Constructing observables}
When constructing observables as outlined in the previous section,
several constraints has to be considered.
\begin{packed_itemize}
\item They should have sensitivity to a range of new physics
  models. In the work we present here, we have concentrated on
  sensitivity to models with right handed currents, but other choices
  are equally possible.
\item In the low \qsq region, where there are only two form factors,
  observables should be constructed where their value cancel out to
  leading order.
\item The effect of $\lqcd/m_\b$ corrections should be demonstrated to
  be small in comparison to expected differences between the SM and
  different NP models.
\item The observable should be invariant under the symmetries of the
  differential distribution described above. Otherwise it would not be
  well defined from an experimental point of view.
\item The experimental resolution from a data set obtainable with \lhcb
  or a super-\B factory should be good enough to offer distinction
  between models.
\end{packed_itemize}

Based on these constraints, a range of \CP-conserving observables have
been designed, designated \AT{i}, with $i=2,3,4,5$.

An example of an observable fulfilling all the criteria is the
asymmetry \AT2, first proposed in~\cite{Kruger:2005ep}, and defined as
\begin{equation}
  \label{eq:AT2general}
  \AT2 =\frac{|A_{\perp}|^2 - |A_\parallel|^2}{|A_\perp|^2 + |A_\parallel|^2},
\end{equation}
where $|A_i|^2 = |A_i^L|^2 + |A_i^R|^2$. It has a simple form, free
from form factor dependencies, in the heavy-quark ($m_{B}\to\infty$)
and large \Kstarzb energy ($E_{K^*}\to\infty$) limits:

\begin{widetext}
\begin{equation}
  \label{eq:AT2generalLEET}
    \AT2 = \frac{2 \left[{\mathrm{Re}}\left(\Cpeff{10} {\Ceff{10}}^{*} \right) + F^2 {\mathrm{Re}}\left(\Cpeff7 {\Ceff7}^{*} \right)  
    + F {\mathrm{Re}}\left(\Cpeff7 {\Ceff9}^{*} \right)\right]}
    {|\Ceff{10}|^2  +  |\Cpeff{10}|^2+ F^2 (|\Ceff7|^2 + |\Cpeff7|^2 ) + |\Ceff9|^2  + 
    2 F {\mathrm{Re}}\left(\Ceff7  {\Ceff9}^{*} \right)},
\end{equation}
\end{widetext}

\begin{figure}
  \centering
  \includegraphics[width=0.98\linewidth]{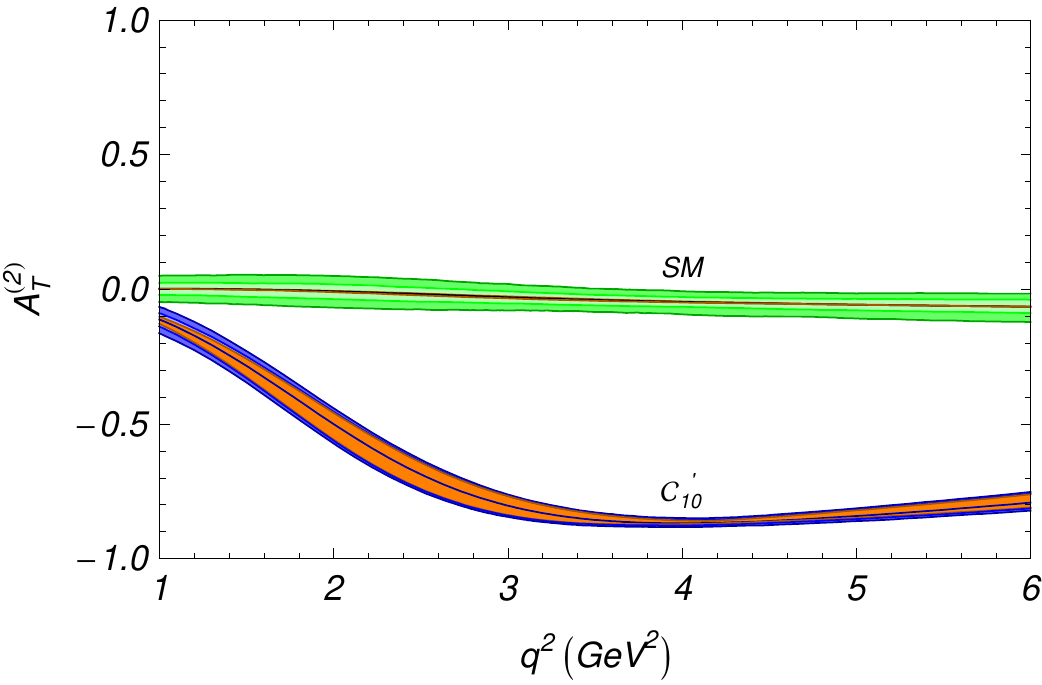}
  \caption{\AT2 in the SM (green) and with NP in $\Cpeff{10}=3 e^{i
      \frac{\pi}{8}}$ (blue); this value is allowed by the model
    independent analysis of \cite{Bobeth:2008ij}. The inner line
    corresponds to the central value of each curve. The dark orange
    bands surrounding it are the NLO results including all
    uncertainties (except for $\Lambda_{\rm QCD}/m_b$). Internal light
    green/blue bands (barely visible) include the estimated
    $\Lambda_{\rm QCD}/m_b$ uncertainty at a $\pm 5\%$ level and the
    external dark green/blue bands correspond to a $\pm 10\%$
    correction for each spin amplitude.}
  \label{fig:AT2Theory}
\end{figure}

\begin{figure}
  \centering
  \includegraphics[width=0.98\linewidth]{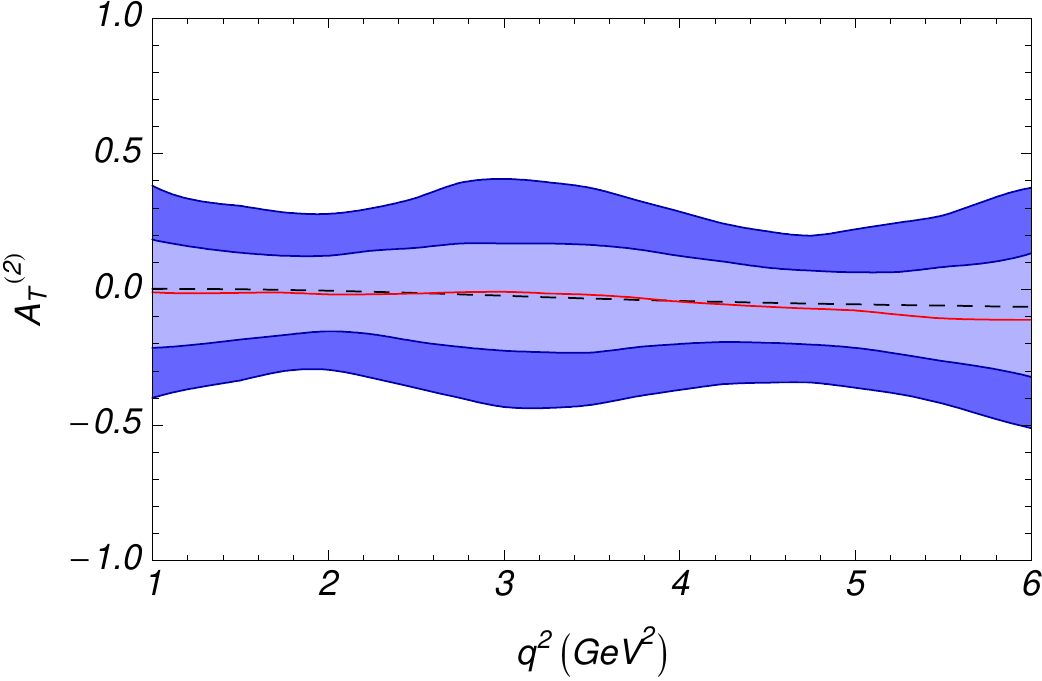}
  \caption{For \AT2 we illustrate the expected statistical
    experimental errors. The inner and outer bands correspond to
    1$\sigma$ and 2$\sigma$ statistical errors with a yield
    corresponding to a 10\invfb data set from \lhcb.}
  \label{fig:AT2Experimental}
\end{figure}

where $F \equiv 2 m_b m_B /q^2$. The sensitivity to the primed Wilson
coefficients corresponding to right handed currents can clearly be
seen from the equation and is illustrated in Fig.~\ref{fig:AT2Theory}
where a NP contribution to $\Cpeff{10}$ is considered. It can also be
seen how the theoretical errors are very small. In
Fig.~\ref{fig:AT2Experimental} the expected experimental resolution
with data from the \lhcb experiment is illustrated. 

\begin{figure}
  \centering
  \includegraphics[width=0.98\linewidth]{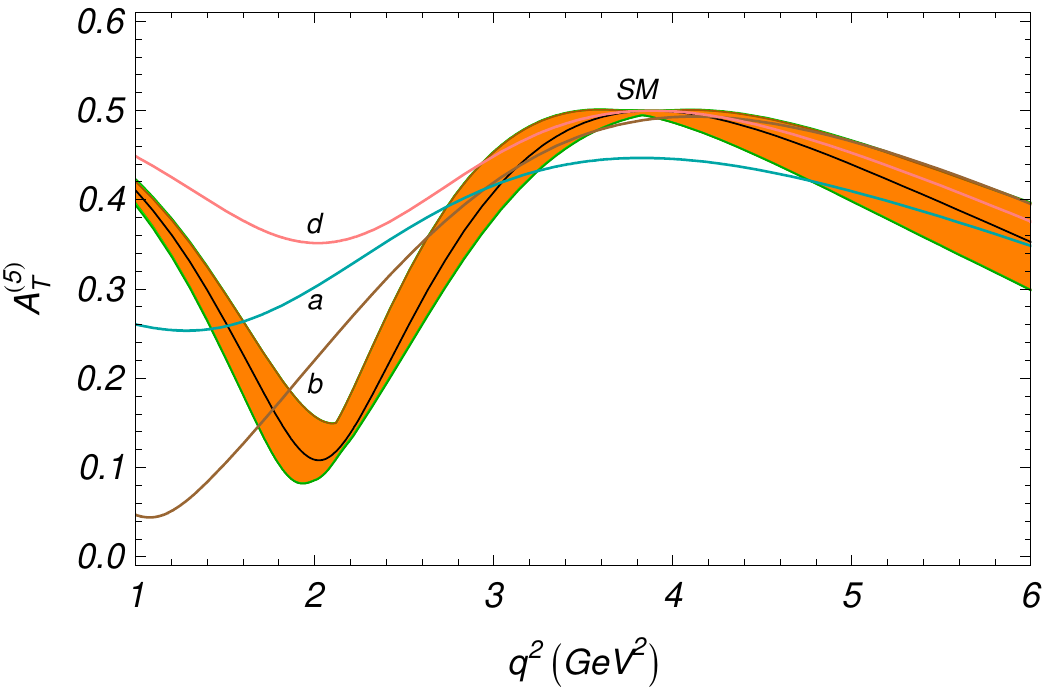}
  \caption{\AT5 in the SM and with NP in both the $\Ceff7$ and $\Cpeff7$
    Wilson coefficients. The cyan line ($a$) corresponds to
    ($\C7^{\scriptscriptstyle{\mathrm{NP}}}$, $\Cp7$) = ($0.26 e^{-i
      \frac{7\pi}{16}}$, $0.2 e^{i \pi}$), the brown line ($b$) to
    ($0.07e^{i \frac{3\pi}{5}}$, $0.3 e^{i \frac{3\pi}{5}}$) and the
    pink line ($d$) to ($0.18 e^{-i \frac{\pi}{2}}$, $0$). The bands
    symbolise the theoretical uncertainty as described in
    Fig.~\protect\ref{fig:AT2Theory}.}
  \label{fig:AT5Theory}
\end{figure}

As another example,
\begin{equation}
 \label{eq:AT5general}
 \AT5 = \frac{\big|\apeL \apaR^* + \apeR^* \apaL\big|}
             {\big|\apeL\big|^2 + \big|\apeR\big|^2 + \big|\apaL\big|^2 + \big|\apaR\big|^2}   \, .
\end{equation}
as defined in~\cite{Egede:2010zc}, has a very different behaviour with
respect to NP contributions, and a comparison of experimental
measurements of \AT2 and \AT5, will be able to provide details of the
underlying theory. The sensitivity to NP of \AT5 is illustrated in
Fig.~\ref{fig:AT5Theory}.

\section{Conclusion}
We have presented how the decay \BdbKsll can provide detailed
knowledge of NP effects in the flavour sector. We developed a method
for constructing observables with specific sensitivity to some types
of NP while, at the same time, keeping theoretical errors from form
factors under control. We demonstrate the possible impact of the
unknown $\Lambda_{\rm QCD}/m_b$ corrections on the NP sensitivity of
the observables. Experimental sensitivity to the observables was
evaluated for datasets corresponding to 10\invfb of data at \lhcb. A
phenomenological analysis of the \AT{i} observables reveals a good
sensitivity to NP including the effects from \CP-violating phases.

\bibliographystyle{JHEP}
\bibliography{BdKstMuMu}

\end{document}